\newfont{\mathbb}{bbold12}
\newfont{\Bbf}{bbold8}
\newcommand{\intR}{\mbox{\Bbf R}} 
\newcommand{\R}{\mathbb R}    
\newcommand{\C}{\mathbb C}    

\newcommand{\Pk}{\mathbb P}

\newcommand{\CP}{\C\Pk^{1}}

\documentclass[12pt]{JHEP}
\usepackage{epsfig,amsfonts,mathrsfs}
\title{Transmogrifying Fuzzy Vortices}
\author{Jeff Murugan and Antony Millner\\
	Department of Mathematics and Applied Mathematics\\
	University of Cape Town\\ Private Bag, Rondebosch 7700\\ 
	South Africa\\
	\email{jeff@hbar.mth.uct.ac.za,amillner@maths.uct.ac.za}}
	
\abstract{We show that the construction of vortex solitons of the 
noncommutative Abelian-Higgs model can be extended to a critically coupled 
gauged linear sigma model with Fayet-Illiopolous D-terms. 
Like its commutative counterpart, this fuzzy linear sigma model has a rich
spectrum of BPS solutions. We offer an explicit construction of the degree$-k$ 
static  semilocal vortex and study in some detail the infinite coupling limit in which it 
descends to a degree$-k$ $\C\Pk^{N}$ instanton. This relation between the fuzzy
vortex and noncommutative lump is used to suggest an interpretation of the
noncommutative sigma model soliton as tilted D-strings stretched between an
NS5-brane and a stack of D3-branes in type IIB superstring theory.} 
  
\begin{document}

\section{Introduction}
\label{Introduction}
A little more than a decade ago, the study of electroweak strings in a
modified Abelian-Higgs theory initiated in \cite{Vachaspati} revealed a
curious new vortex solution. As the story goes, vortices are indeed
enigmatic objects \cite{Tong} and the {\it semilocal vortices} found in
\cite{Vachaspati} are no exception. Firstly, standard lore holds that
a non-simply-connected vacuum manifold is a necessary condition for the
existence of stable, finite energy cosmic string solutions.
If this is anything to go by, the very existence of these semilocal 
vortices should be called into question since the vacuum manifold of 
the modified Abelian-Higgs theory is $S^{3}$. 
Yet exist they do. Consequently, a more consistent condition was offered 
in \cite{Hindmarsh93}. 
Semilocal vortices (actually, this holds for other defects as well) 
form in theories exhibiting spontaneous symmetry breaking and whose 
vacuum manifold is fibred by the action of the gauge group in some 
non-trivial way. In this same work it was realised also that the low
momentum dynamics of these vortices bear a striking resemblance to the 
$2-$dimensional lump solutions of the $\C\Pk^{N}$ nonlinear sigma model.
Since then, this similarity between the modified Abelian-Higgs theory (a.k.a
gauged linear sigma model) and the $\C\Pk^{N}$ (or, more generally,
Grassmannian) sigma model has found itself the subject of much attention
\cite{Hanany-Tong,Schroers,Witten93}. Nevertheless, much of what is 
known about the semilocal vortex is only asymptotic. Even its descent to the
lump in the infinite coupling limit is only exact at spatial infinity and
suffers Skyrme term corrections at smaller radial distances. This is the
allure and frustration of vortices; as simple as their 
defining equations seem, they are also remarkably unyielding.\\

\noindent
Until a short time ago, the only avenue toward tractable vortex equations 
was a curvature deformation of the background space in which the vortices 
live \cite{Strachan,Witten77}. These are, of course, not without their own
puzzles. The recent renaissance in noncommutative geometry (due in no
small part to the seminal work of \cite{SeibergWitten}) offers new
recourse. {\it Fuzzy deformations} of the background space have, in 
only a few years, not only yielded a wealth of new solitonic solutions but
also several new insights into old solutions to a host of field theories 
(see \cite{Douglas-Nekrasov,Harvey1,Szabo} for excellent reviews). The {\it
noncommutative} Abelian-Higgs model, for example, exhibits exact vortex
solutions \cite{DongsuBak,Bak,Lozano1,Wadia} whose moduli space metric can
be computed explicitly in the large noncommutativity limit \cite{Tong}.\\ 

\noindent
In this work we extend this idea to the $(2+1)-$dimensional, critically
coupled, gauged linear sigma model with an $N+1$ component Higgs field. 
The BPS spectrum of the resulting fuzzy theory is studied and, like its 
commutative counterpart, shown to have quite rich structure. In particular,
we use an extension of the computational technique of \cite{Lozano1} to
explicitly construct a family of exact semilocal vortices. As expected, our
family contains the Abelian-Higgs vortices of 
\cite{DongsuBak,Lozano1,Wadia} as well as the fluxons of 
\cite{HarveyKrausLarsen} as special cases. As suggested by the title, the
metamorphosis of the semilocal vortex is of central importance in this
paper. By turning up the gauge coupling, we demonstrate conclusively, at the
level of the solutions, the descent of the semilocal vortex into the
instanton solution of the fuzzy $\C\Pk^{N}$ model of the same degree.
Interestingly, unlike the commutative case, this ``transmogrification" of
the vortex is exact at a certain point in the parameter space of the theory.
Finally, we turn our attention towards the physical\footnote{By which we 
mean `stringy'.} interpretation of the $k-$lump solution of the 
noncommutative $\C\Pk^{N}$ model of \cite{LeeLeeYang}. Without much
additional work, the brane configuration in type II-B string theory that
realises the fuzzy lump may be read off from the construction of
\cite{Hanany-Tong} as tilted $D-$strings suspended between an $NS5-$ 
and $D3-$brane. We conclude, as is conventional, with the conclusions. 
      
\section{The Gauged Linear Sigma Model}
\subsection{Definitions}
\noindent
Among the many extensions to the Abelian-Higgs model, one of the most natural
is the gauged linear sigma model with Fayet-Illiopolous D-terms
\cite{Schroers,Witten93}. This is certainly true if the aim is the construction 
of a model that supports solitonic excitations saturating BPS-like bounds.  With
its $\C^{N+1}$-valued scalar fields and $U(1)^{N+1}$ gauge symmetry, the linear 
sigma model is a natural springboard for our discussion of the relation 
between noncommutative semilocal vortices, fuzzy sigma model lumps and the 
braney systems they are associated with. To this end then it will prove useful
to briefly review some of the ideas and notation used to extract the 
vortex excitations from the solution spectrum of the semilocal model. Following
\cite{Schroers} we write the linear sigma model action as    
\begin{eqnarray}
  S_{SL} &=& -\int_{\intR^{(1,2)}}d^{3}x
  {\Bigl[}(D_{\mu}\Phi)(D^{\mu}\Phi)^{\dagger} + 
  \sum_{a=1}^{N+1}\frac{1}{4e_{a}^{2}}(F^{a}{}_{\mu\nu})^{2}\nonumber\\ 
  &+& 
  \sum_{a=1}^{N}\frac{e_{a}^{2}}{2}(R_{a} - \Phi\tau_{a}\Phi^{\dagger})^{2}{\Bigl]}
  \label{Semilocal Action}
\end{eqnarray}
The dynamical degrees of freedom in this model are encoded in a $\C^{N+1}$-valued spacetime
scalar $\Phi = (\phi_{1},\ldots,\phi_{N+1})$ and the $N+1$ $U(1)$-valued 
connection forms $A^{a} = A^{a}{}_{\mu}dx^{\mu}$ with associated 
curvature $2$-forms $F^{a} = dA^{a}$. The $\tau_{a}$ are the $N+1$ generators of 
$U(1)^{N+1}$. The gauge covariant derivative we will take as 
\begin{eqnarray}
  D := d - i\sum_{a=1}^{N+1}\tau_{a}A^{a} 
\end{eqnarray}
There are two sets of parameters in the theory; the $N+1$ coupling constants
$e_{a}$ of dimension $(mass)^{1/2}$ and $N+1$ Fayet-Illiopolous (FI) parameters 
$R_{a}$ - effectively the vacuum expectation values of the components of $\Phi$.
Without loss of generality (and because we can always re-scale the fields to
absorb them anyway) we set the latter to unity. The coupling constants we 
retain because they control the energy scales of the model.\\      

\noindent
In the temporal gauge, the static energy corresponding to the action 
(\ref{Semilocal Action}) is 
\begin{eqnarray}
  E = \int_{\intR^{2}}d^{2}x\,
  {\Bigl[}(D_{i}\Phi)(D_{i}\Phi)^{\dagger} +
  \sum_{a=1}^{N+1}\frac{1}{4e_{a}^{2}}(F^{a}{}_{ij})^{2} + 
  \sum_{a=1}^{N+1}\frac{e_{a}^{2}}{2}(\Phi\tau_{a}\Phi^{\dagger} - 1)^{2}{\Bigr]}
  \label{SemilocalStaticEnergy}
\end{eqnarray}
For instance, in the case $N=1$, following \cite{Schroers} 
the energy functional becomes (in exhaustive detail)
\begin{eqnarray}
  E = \int_{\intR^{2}}d^{2}x\,&{\Bigl[}&(D_{1}\Phi)
  (D_{1}\Phi)^{\dagger} +
  (D_{2}\Phi)(D_{2}\Phi)^{\dagger} +
  \frac{1}{4e_{1}^{2}}(F_{ij})^{2} + 
  \frac{1}{4e_{2}^{2}}(G_{ij})^{2}\nonumber\\
  &+& \frac{e_{1}^{2}}{2}(\Phi\tau_{1}\Phi^{\dagger} - 1)^{2} 
  + \frac{e_{2}^{2}}{2}(\Phi\tau_{2}\Phi^{\dagger} - 1)^{2}
  {\Bigr]}
  \label{StaticEnergy2}
\end{eqnarray}
where the $GL(2,\R)$-valued connections $A = \tau_{1}A^{1}$ 
and $B = \tau_{2}A^{2}$ are associated to the curvature forms 
$F = dA$ and $G = dB$ respectively. For our purposes, it will 
suffice to turn off $B$ and $e_{2}$ and take $\tau_{1}={\bf 1}_{2}$ giving
\begin{eqnarray}
  E = \int_{\intR^{2}}d^{2}x\,
  {\Bigl[}(D_{1}\Phi)(D_{1}\Phi)^{\dagger} +
  (D_{2}\Phi)(D_{2}\Phi)^{\dagger} +
  \frac{1}{2e_{1}^{2}}(F_{ij})^{2} + 
  \frac{e_{1}^{2}}{4}(\Phi\Phi^{\dagger} - 1)^{2} 
  {\Bigr]}
  \label{StaticEnergy3}
\end{eqnarray} 
Even under such restricted circumstances,
the resulting linear sigma model is still remarkably rich 
\cite{Schroers,Witten93}, exhibiting a wealth of solitonic structure and
enjoying intimate relations with nonlinear sigma models on 
toric varieties.\\ 

\noindent
In what follows, it will prove convenient to rewrite the 
energy in terms of the
complex coordinates $z:=(x^{1}+ix^{2})/\sqrt{2}$ and ${\bar{z}} 
:=(x^{1}-ix^{2})/\sqrt{2}$. This particular normalization means that
\begin{eqnarray}
  \partial_{z}:=\frac{\partial}{\partial z} = \frac{1}{\sqrt{2}}
  {\Bigl(}\partial_{1} - i\partial_{2}{\Bigr)}\qquad
  \partial_{\bar{z}}:=\frac{\partial}{\partial \bar{z}} = \frac{1}{\sqrt{2}}
  {\Bigl(}\partial_{1} + i\partial_{2}{\Bigr)}
\end{eqnarray} 
This in turn induces a complexification of the gauge covariant derivative so that
\begin{eqnarray}
  D_{z}:= \frac{1}{\sqrt{2}}
  {\Bigl(}D_{1} - iD_{2}{\Bigr)}\qquad
  D_{\bar{z}}:= \frac{1}{\sqrt{2}}
  {\Bigl(}D_{1} + iD_{2}{\Bigr)}.
\end{eqnarray}
when $A_{z}:=(A_{1}-iA_{2})/\sqrt{2}$ and 
$A_{\bar{z}}:=(A_{1}+iA_{2})/\sqrt{2}$. These are of course now
$GL(2,\C)$-valued objects. With these definitions, 
\begin{eqnarray}
 E = \int_{\C}d^{2}z\,{\Bigl[}(D_{z}\Phi)(D_{z}\Phi)^{\dagger} +
  (D_{\bar{z}}\Phi)(D_{\bar{z}}\Phi)^{\dagger} +
  \frac{1}{2e^{2}}(F_{12})^{2} + 
  \frac{e^{2}}{2}(\Phi\Phi^{\dagger} - 1)^{2} 
  {\Bigr]}
  \label{StaticEnergy4}
\end{eqnarray}

\subsection{Solitons on the Plane}
\noindent
To see the emergence of the semilocal vortex in the spectrum of the gauged
linear sigma model, the usual method of ``completing the square" in
the energy functional may be followed. After some straightforward manipulations,
(\ref{StaticEnergy4}) may be put into the form
\begin{eqnarray}
 E &=& \int_{\C}d^{2}z\,{\Bigl[} 
 2(D_{\bar{z}}\Phi)(D_{\bar{z}}\Phi)^{\dagger} +
 \frac{1}{2e^{2}}{\Bigl|}F_{12} + e^{2}(\Phi\Phi^{\dagger} - 1){\Bigr|}^{2} 
 {\Bigr]}\nonumber\\ 
 &+& \int_{\C}d^{2}z\,T + 
 \underbrace{\int_{\C}d^{2}z\,F_{12}}_{2\pi k}
 \label{StaticEnergyBPSForm}
\end{eqnarray}
where $T = \partial_{\bar{z}}(\Phi D_{z}\Phi^{\dagger}) -
\partial_{z}(\Phi D_{\bar{z}}\Phi^{\dagger})$. As such, the second to last term is
a total derivative whose integral vanishes. Consequently, a nonvanishing lower
bound of $E \geq 2\pi k$ is established
on finite energy field configurations. As usual, the bound saturates when the
first order system
\begin{eqnarray}
 D_{\bar{z}}\Phi &=& 0\nonumber\\
 F_{12} &=& {e^{2}}(\Phi\Phi^{\dagger} - 1)\nonumber\\
 \int_{\C}d^{2}z\,F_{12} &=& 2\pi k
 \label{CommutativeBPS}
\end{eqnarray}
is satisfied. The first of these is, of course, really two equations, one for
each component of the $\C^{2}$-valued field $\Phi$. The equations
in (\ref{CommutativeBPS}) form a closed system whose solutions are precisely the
semilocal vortices of \cite{Hindmarsh92,Schroers,Vachaspati}.\\ 

\noindent
Although such solitonic solutions are
vortex-like in many respects, a little analysis soon reveals that their
asymptotic behavior is very different from the exponential falloff of
Abelian-Higgs vortices \cite{Hindmarsh92,Hindmarsh93}. 
In fact the fields of the semilocal model exhibit a
distinctive power-law behavior at spatial infinity, a symptom of the fact that the
width of the flux tube is an arbitrary parameter of the theory. This should be
contrasted with Abelian-Higgs vortices where the width is controlled by the
Compton wavelength of the gauge boson. In this sense, these vortex solutions are
rather reminiscent of $\C\Pk^{N}$ instantons. This is no mere coincidence. In
fact, the correspondence can be made precise in the large coupling
limit in which the semilocal vortices of the $U(1)^{N+1}$-gauged linear sigma
model descend to the instanton solutions of a $\C\Pk^{N}$ nonlinear sigma
model \cite{Schroers,Witten93}. While this is quite clear at the levels of the
action and equations of motion, its realization at the level of the solutions
is marginally obscured by the fact that only the asymptotic forms of the vortex
solutions are known to exist. This is not unlike the situation with the
conventional Nielsen-Olesen vortex. However this particular hurdle was recently
surmounted in \cite{DongsuBak,Bak,Lozano1,Wadia} where a noncommutative
deformation of the two-dimensional configuration space of the Abelian-Higgs
model allows for the construction of {\it exact vortex solutions}. The fact 
that the noncommutative version of the theory seems so much richer than its 
commutative counterpart is by now not surprising \cite{GMS,GHS}. 
It would seem then, that a noncommutative deformation of the base space of 
the two-dimensional gauged linear sigma model might offer, if nothing else, 
an interesting avenue to explore the construction of exact semilocal 
vortices.    

\section{And then everything became fuzzy...}
\noindent
Conventionally a noncommutative deformation of the two-dimensional 
configuration space 
is imposed by a Moyal-deformation of the algebra of functions over $\R^{2}$ and
implemented by replacing ordinary pointwise multiplication of functions by
Moyal $*$-multiplication. Consequently\footnote{Our conventions for the 
noncommutative theory follow \cite{Murugan-Adams}}, coordinates on 
the noncommutative plane $\R^{2}_{\vartheta}$ satisfy the Heisenberg algebra
\begin{eqnarray}
 [x^{i},x^{j}] = x^{i}*x^{j} - x^{j}*x^{i} = i\theta^{ij}
\end{eqnarray}
where $\theta^{ij}=\vartheta\epsilon^{ij}$ is a nondegenerate, antisymmetric 
matrix of constants and $\vartheta$ is a positive deformation parameter of 
dimension $(mass)^{2}$. In terms of the complex coordinates on 
$\R^{2}_{\vartheta}$, the commutator $[z,\bar{z}]=\vartheta$ is easily seen to be isomorphic to the 
algebra of annihilation and creation operators 
for the simple harmonic oscillator. Thus a function on the  
noncommutative space may be associated, via a Weyl transform \cite{Harvey1}, 
to an operator acting on an 
auxiliary one-particle Hilbert space ${\cal H} = \bigoplus_{n}\C |n\rangle$ 
built from harmonic oscillator eigenstates. After a mild redefinition of 
$\widehat{z}=\sqrt{\vartheta}\widehat{a}$ and 
$\widehat{\bar{z}}=\sqrt{\vartheta}\widehat{a}^{\dagger}$, the
action of the coordinate operators on the basis states is given by
\begin{eqnarray}
 \widehat{a}|n\rangle &=& \sqrt{n}|n-1\rangle\nonumber\\
 \widehat{a}^{\dagger}|n\rangle &=& \sqrt{n+1}|n+1\rangle,
\end{eqnarray}
with the vacuum $|0\rangle$ defined by $\widehat{a}|0\rangle = 0$.
Further, under the Weyl map
\begin{eqnarray}
 \partial_{z}&\rightarrow& 
 -\frac{1}{\sqrt{\vartheta}}[\widehat{a}^{\dagger},\cdot]\\
 \int_{\C_{\vartheta}}\,d^{2}z\,f(z,\bar{z})&\rightarrow& 
 2\pi\vartheta\,{\rm Tr}_{\cal H}\,\widehat{O}_{f}(\widehat{z},\widehat{\bar{z}})
\end{eqnarray} 
with an analogous expression for $\partial_{\bar{z}}$.

\subsection{The Noncommutative Semilocal Model}
With this prescription at hand, the noncommutative semilocal energy functional
(\ref{StaticEnergy4}) can be written as
\begin{eqnarray}
 E_{\vartheta} 
 &=& 2\pi\vartheta {\rm Tr}_{\cal H}
 {\Bigl[}(\widehat{D_{z}}\widehat{\Phi})(\widehat{D_{z}}\widehat{\Phi})^{\dagger} + 
 (\widehat{D_{\bar{z}}}\widehat{\Phi})(\widehat{D_{\bar{z}}}
 \widehat{\Phi})^{\dagger} + 
 \frac{1}{2e^{2}}\widehat{F_{12}}^{2} + 
 \frac{e^{2}}{2}(\widehat{\Phi}\widehat{\Phi}^{\dagger} - 1)^{2}
 {\Bigr]}
 \label{NCStaticEnergy1}
\end{eqnarray} 
where, now $\widehat{D_{z}}\widehat{\Phi} =
(\widehat{\Phi}\widehat{a}^{\dagger} +
\widehat{C}^{\dagger}\widehat{\Phi})/\sqrt{\vartheta}$, 
$\widehat{D_{\bar{z}}}\widehat{\Phi} =
-(\widehat{\Phi}\widehat{a} + \widehat{C}\widehat{\Phi})/\sqrt{\vartheta}$ and 
the gauge field is parameterized as
$\widehat{A}_{z}=(i/\sqrt{\vartheta})(\widehat{a}^{\dagger} + 
\widehat{C}^{\dagger})$. As in the commutative case, this can be massaged 
into a Bogomol'nyi form which is saturated when the BPS equations 
\begin{eqnarray}
  \widehat{\Phi}\widehat{a} + \widehat{C}\widehat{\Phi} &=& 0\nonumber\\
  1+ {\Bigl[}\widehat{C}^{\dagger},\widehat{C}{\Bigr]} &=&
  \vartheta e^{2}(\widehat{\Phi}\widehat{\Phi}^{\dagger}-1)\nonumber\\
  \rm{Tr}_{\cal H}{\Bigl(}1 + 
  {\Bigl[}\widehat{C}^{\dagger},\widehat{C}{\Bigr]}{\Bigr)} &=& -k
  \label{NCBPS-equations}
\end{eqnarray}
are satisfied. As in the commutative case, this is a system of three first 
order equations, subject to the flux constraint. Solutions of this system 
will be the noncommutative generalizations of the semilocal vortex of 
\cite{Hindmarsh93}. In the spirit of \cite{Bak, Lozano1}, we begin with an 
{\it ansatz} for the Higgs doublet and the gauge field. 
To this end the most general vortex-like solution of the 
BPS equations which maintain the cylindrical symmetry is of the
form\footnote{The generalization to an $N+1$ component Higgs field is quite
straightforward so we persist in restricting our attention to the $N=1$ case 
for the moment.}
\begin{eqnarray}
  \widehat{\Phi} = \widehat{\phi_{1}}\otimes \langle\rm{I}| + 
  \widehat{\phi_{2}}\otimes \langle\rm{I\!I}|
  \label{Ansatz1} 
\end{eqnarray}
where $\langle\rm{I}| = (1,0)$, $\langle\rm{I\!I}| = (0,1)$ and   
\begin{eqnarray}
  \widehat{\phi_{i}} &=& \sum_{m=0}^{\infty}\,f^{(i)}_{m}|m\rangle\langle 
  m+q^{(i)}|
  \label{Ansatz2}
\end{eqnarray}
where $\{q^{(1)},q^{(2)}\}$ is a set of integers related to the topological
charge and, respectively, angular momentum quantum number of the vortex as we 
show below. For the $U(1)$ gauge field we take the cylindrically symmetric 
ansatz 
\begin{eqnarray}
  \widehat{C} = \sum_{m=0}^{\infty} g_{m}|m\rangle\langle m+1|.
\end{eqnarray}
Without loss of generality all coefficients are taken to be real. 
The construction of exact vortex solutions to the semilocal model now hinges on
determining the various coefficients in the above {\it ansatze} that satisfy
the appropriate boundary conditions. In terms of the coefficients
$f_{n}\equiv f_{n}^{(1)}$ and $h_{n}\equiv f_{n}^{(2)}$, the first of 
eqs.(\ref{NCBPS-equations}) become
\begin{eqnarray}
  f_{m}\sqrt{m+q+1} + g_{m}f_{m+1}&=& 0\nonumber\\
  h_{m}\sqrt{m+1}+ g_{m}h_{m+1}&=& 0 
  \label{Recurrence-BPS1}
\end{eqnarray}
with the choice $q^{(1)}=q$ and $q^{(2)}=0$. An explanation for this
will follow in due course. For the moment though, notice that 
eqs.(\ref{Recurrence-BPS1}) mean that the coefficients of each of the
components of the Higgs doublet are not independent. Indeed
\begin{eqnarray}
  h_{m+1} = \sqrt{\frac{m+1}{m+q+1}}{\Bigl(}\frac{f_{m+1}}{f_{m}}
  {\Bigr)}h_{m}.
\end{eqnarray}
This is a simple recurrence relation which is easily solved for $h_{m}$ to give
\begin{eqnarray}
  h_{m} = \sqrt{\frac{m!q!}{(m+q)!}}\kappa f_{m}
  \label{Phi2-coefficients}
\end{eqnarray}
with $\kappa = h_{0}/f_{0}$ determining the
relationship between the initial conditions of each coefficient sequence.
With the convenient definitions of $Q_{n}\equiv f_{n}^{2}$ and $P_{n}\equiv
h_{n}^{2}$, this may be combined with the second of the BPS equations to give
\begin{eqnarray}
 Q_{1} &=& \frac{(q+1)Q_{0}}{1+\gamma -\gamma(1+\kappa^{2})Q_{0}}\nonumber\\
 Q_{m+1} &=& \frac{(m+q+1)Q_{m}^{2}}{Q_{m}+(m+q)Q_{m-1} - \gamma
 Q_{m}{\Bigl[}{\Bigl(}1+\frac{m!q!}{(m+q)!}\kappa^{2}{\Bigr)}Q_{m} 
 - 1{\Bigr]}}\quad m>0
 \label{Phi1-coefficients}
\end{eqnarray}
where, following \cite{Tong} the 
dimensionless combination of $\vartheta e^{2}$ is hereafter christened $\gamma$.
In principle then, the noncommutative vortex solution of the critically 
coupled linear sigma model may be determined by solving the recurrence 
relation (\ref{Phi1-coefficients}) and consequently (\ref{Phi2-coefficients}) 
subject to the ``boundary conditions" 
$f_{n}\rightarrow 1$, $h_{n}\rightarrow 0$ as $n\rightarrow \infty$. Well,
almost. The attentive reader would of course have noticed that there is 
still the matter of the arbitrary integer $q$. Fortunately, there is also the
third of the BPS equations, the flux constraint. Using 
(\ref{Recurrence-BPS1}) and the {\it ansatz} for the gauge field it may be
shown that
\begin{eqnarray}
  {\rm Tr}_{\cal H}{\Bigl(}1 + [\widehat{C}^{\dagger},\widehat{C}]{\Bigr)} 
  &=& {\rm Tr}_{\cal
  H}{\Bigl(}\sum_{m=0}^{\infty}{\Bigl(}1+g_{m-1}^{2}
  -g_{m}^{2}{\Bigr)}|m \rangle\langle m|{\Bigr)}\nonumber\\
  &=& \sum_{l,m=0}^{\infty}{\Bigl(}1+g_{m-1}^{2}
  -g_{m}^{2}{\Bigr)}\langle l|m\rangle\langle m|l\rangle\nonumber\\
  &=& \lim_{M\rightarrow\infty}{\Bigl[}M+1 - 
  (M+q+1)\frac{Q_{M}}{Q_{M+1}}{\Bigr]}.
  \label{FluxConstraint}
\end{eqnarray}
In the last step, the cutoff of \cite{Wadia} was employed to regulate the
trace. In the large $M$ limit, the convergence of the coefficient sequence 
means that the ratio of successive $Q$'s approaches unity. Consequently, the
flux constraint equation implies that $q = k$. Indeed, a quick comparison 
with the analogous commutative result confirms that this is the only
physically meaningful conclusion; the index of $\phi_{1}$ is equal to the 
topological number of the vortex. Interestingly enough, choosing
$q^{(2)}\neq 0$ does not affect this conclusion. Again, this is not 
altogether unexpected since $q^{(2)}$ is just the angular momentum 
quantum number of the vortex \cite{Hindmarsh92}. Convergence of the 
coefficient sequence for $\phi_{2}$ bounds the angular momentum quantum number 
to the range $0\leq q^{(2)}< k$. However, since none of the arguments presented
here depends essentially on $q^{(2)}$ we can, without any loss of generality, set
$q^{(2)}=0$. It is also worth noting that when $q=0$, $h_{m} = \kappa f_{m}$
and both boundary
conditions can only be simultaneously satisfied if $h_{m}\equiv 0$ which
reduces to the $k=0$ vortex of the noncommutative Abelian-Higgs model 
\cite{DongsuBak,Bak,Lozano1,Tong,Wadia}. Instead of solving 
eqs.(\ref{Phi1-coefficients}) in full generality, it is perhaps more
illuminating to focus on a few examples.
     
\subsection{Examples}
\begin{enumerate}
 \item
   To begin with we consider the case $h_{m} = 0$ for all $m$. In this case
   the Higgs doublet $\Phi = (\phi_{1},0)$ satisfies exactly the equations 
   of motion of the noncommutative Abelian-Higgs model and it is quite easy
   to check that the solutions of (\ref{Phi1-coefficients}) reduce to the
   degree$-k$ vortices found in \cite{Lozano1} for which
   \begin{eqnarray}
     Q_{1} &=& \frac{(q+1)Q_{0}}{1+
     \gamma(1 - Q_{0})}\nonumber\\
     Q_{m+1} &=& \frac{(m+q+1)Q_{m}^{2}}{Q_{m}+(m+q)Q_{m-1} - \gamma
     Q_{m}{\Bigl(}Q_{m} - 1{\Bigr)}}\qquad m>0
     \label{Phi1-coefficients-NCAH}
   \end{eqnarray}
   This set of equations has been studied extensively and numerically shown
   to exhibit regular vortex solutions with $+k$ units of magnetic flux for
   a large $\gamma$ range. In particular, for small $\vartheta$ (and
   consequently $\gamma$) the regular commutative Neilsen-Olesen vortex
   solutions of \cite{deVega,Nielsen-Olesen} are obtained. In addition, an
   obvious solution to (\ref{Phi1-coefficients-NCAH}) that satisfies the
   boundary conditions of the semilocal model is $Q_{m} \equiv 1$. 
   As noted in \cite{Lozano1}, these are exactly the fluxon 
   solutions of \cite{HarveyKrausLarsen}.
 \item
   Moving on now to the more interesting case of non-vanishing $h_{m}$, 
   it will suffice to restrict our attention to $k=1$ for which 
   the BPS recurrence relations become
   \begin{eqnarray}
     P_{m} &=& \frac{\kappa^{2}}{m+1}Q_{m}\nonumber\\
     Q_{1} &=& \frac{2Q_{0}}{1+\gamma -\gamma(1+
     \kappa^{2})Q_{0}}\nonumber\\
     Q_{m+1} &=& \frac{(m+2)Q_{m}^{2}}{Q_{m}+(m+1)Q_{m-1} - \gamma
     Q_{m}{\Bigl[}{\Bigl(}1+\frac{\kappa^{2}}{m+1}{\Bigr)}Q_{m} 
     - 1{\Bigr]}}\qquad m>0.\qquad{}
     \label{Phi1-coefficients-k=1}
   \end{eqnarray}
   The vortex solutions of the noncommutative semilocal model are constructed by
   solving eqs.(\ref{Phi1-coefficients-k=1}) subject to the convergence 
   constraint $(P_{m},Q_{m})\rightarrow (0,1)$ as $n\rightarrow\infty$. From the
   first of these it is clear that when the $Q_{m}$ sequence converges and
   $\kappa$ is of order unity, $P_{m}\sim 1/m$ for large $m$. Again, 
   this remains true for any fixed value of the angular momentum 
   quantum number. We solve the above system numerically using a double
   precision, split-step 
   shooting algorithm. At first glance, the shooting-parameter space looks 
   to be two-dimensional (corresponding to the different values of the pair
   $(P_{0},Q_{0})$) but a prescient choice of $\kappa^{2} = 1/\vartheta$ fixes one
   of these parameters in terms of the other and reduces
   the dimension to one. With the initial value $Q_{0}$ as the shooting parameter, we
   solve (\ref{Phi1-coefficients-k=1}) for various values of $\gamma$ and
   tabulate our results below.
   \begin{center}
   \begin{tabular}{|l|l|l|l|}
     \hline
     $\vartheta$ & $e^{2}$ & $\gamma$ & $Q_{0}$\\
     \hline
     0.2 & 1  & 0.2 & 0.099732894\\
     0.2 & 4  & 0.8 & 0.140471163\\
     0.2 & 16 & 3.2 & 0.158732886334\\
     0.2 & 36 & 7.2 & 0.16297094403243935\\
     0.5 & 1  & 0.5 & 0.215729007\\
     0.5 & 4  & 2   & 0.2895665841653\\
     0.5 & 16 & 8   & 0.32043540606185\\
     0.5 & 36 & 18  & 0.32737242959721649\\
     \hline
   \end{tabular}
   \end{center}
   Each of these initial values for the $Q_{m}$ results in a coefficient
   sequence that converges (with varying degrees of accuracy) to one. Once
   determined, the $P_{m}$ and $Q_{m}$ may then be used to compute other
   characteristic quantities associated with the semilocal vortex. 
   For example, the magnetic field of the semilocal vortex may easily be 
   computed as
   \begin{eqnarray}
     \widehat{B} = \frac{\gamma}{\vartheta}\sum_{n=0}^{\infty}
     {\Bigl[}1 - {\Bigl(}\frac{n+1+\vartheta^{-1}}{n+1}{\Bigr)}Q_{n}{\Bigr]}
     |n\rangle\langle n|.
     \label{BPS-magnetic-field}
   \end{eqnarray}
   Substituting this, together with the covariant derivative
   \begin{eqnarray}
     \widehat{D}_{z}\widehat{\Phi} &=& 
     \sum_{m=0}^{\infty}\frac{1}{\sqrt{\vartheta}}
     {\Bigl(}f_{m}\sqrt{m+1} + g_{m-1}f_{m-1}{\Bigr)}|m\rangle\langle m|\otimes
     \langle {\rm I}|\nonumber\\ 
     &+& \sum_{m=0}^{\infty}\frac{1}{\sqrt{\vartheta}}
     {\Bigl(}h_{m+1}\sqrt{m+1} + g_{m}f_{m}{\Bigr)}|m+1\rangle\langle m|\otimes
     \langle {\rm I\!I}|
   \end{eqnarray}
   into eq.(\ref{NCStaticEnergy1}) allows for the energy density of the vortex to
   be computed quite straightforwardly as
   \begin{eqnarray} 
     \widehat{{\cal E}} 
     &=& \frac{1}{\vartheta}\sum_{m=0}^{\infty}
     {\Bigl[}\frac{m+1}{Q_{m}}(Q_{m}-Q_{m-1})^{2} + 
     \frac{m}{P_{m}}(P_{m}-P_{m-1})^{2}\nonumber\\
     &+& 
     \gamma{\Bigl(}1 + (\frac{m+1+\vartheta^{-1}}{m+1})Q_{m}{\Bigr)}^{2}
     {\Bigr]}|m\rangle\langle m|.
     \label{1-vortex-energy-density}   
   \end{eqnarray}
   It may be verified numerically that up to the first few hundred terms the 
   above expression for the energy density sums to $1/(2\pi\vartheta)$ to within 
   a few percent as is expected for the $1-$vortex solution. To make contact with
   the primary aim of this paper, it will be convenient to visualize the profile
   of the vortex, especially as $\gamma$ is turned up. However, both 
   eqs.(\ref{BPS-magnetic-field}) and (\ref{1-vortex-energy-density}) are
   Fock space representations. Fortunately, these can be turned into
   (noncommutative) coordinate space representations relatively easily with the
   inverse Weyl map under which the projector $| n\rangle\langle n|\mapsto
   (-1)^{n}\exp{(-r^{2}/\vartheta)}L_{n}(-2r^{2}/\vartheta)$ where $L_{n}$ is the
   $n$'th Laguerre polynomial. In fig.1 we plot the magnetic field as a function
   of $r$ for various values of the dimensionless parameter $\gamma$. Fig.2.
   contains a series of snapshots of the energy profile of the vortex as gamma
   increases from $0.2$ to $28.8$.
\end{enumerate}
\begin{figure}{\epsfig{file=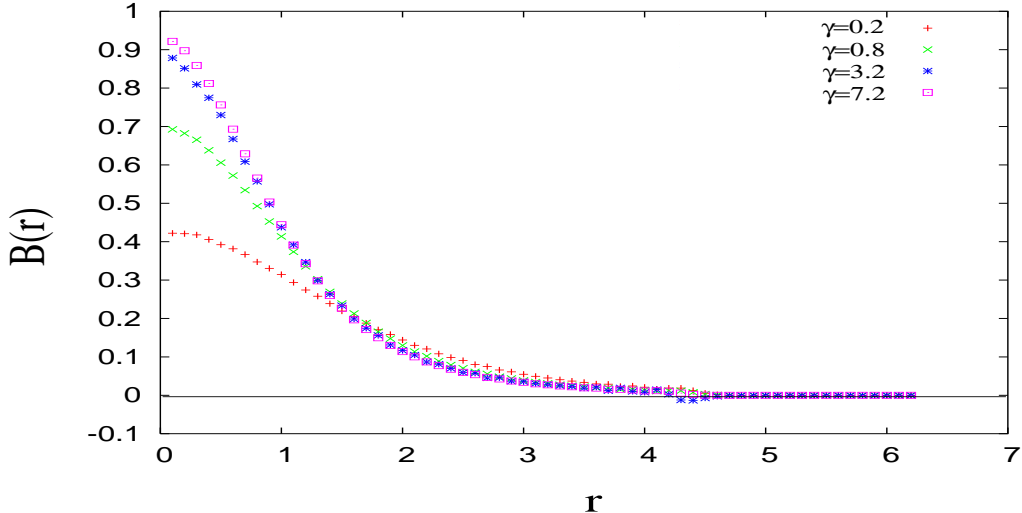,width=14cm,height=7cm}         
   \caption{The magnetic field trapped in the vortex core for varying
   $\gamma$}}
\end{figure}
\section{The large coupling limit}
Having presented a general algorithm for the construction of degree$-k$ 
semilocal vortex solutions of the gauged noncommutative linear sigma model 
and explicitly constructed the $1-$vortex solution we proceed now to study 
one of the more interesting limits of the semilocal model: its large coupling 
limit. At the level of the action (\ref{NCStaticEnergy1}), the $e^{2}
\rightarrow\infty$ limit decouples the gauge field dynamics and any finite 
energy static solution has
\begin{eqnarray}
 E
 &=& 2\pi\vartheta {\rm Tr}_{\cal H}
 {\Bigl[}
 (\widehat{D}_{z}\widehat{\Phi})(\widehat{D}_{z}\widehat{\Phi})^{\dagger} + 
 (\widehat{D}_{\bar{z}}\widehat{\Phi})(\widehat{D}_{\bar{z}}
 \widehat{\Phi})^{\dagger}
 {\Bigr]} 
 \label{Large-Coupling-Energy}
\end{eqnarray}  
subject to the constraint $\widehat{\Phi}\widehat{\Phi}^{\dagger}=1$. 
In this limit, the gauge field
is relegated to an auxiliary field, completely determined by $\Phi$. 
Recalling that $\Phi$ is an $(n+1)-$component complex vector leads to the
conclusion that this is, of course, nothing but the noncommutative version of 
the $\C\Pk^{N}$ sigma model. At the level of the action this observation is 
certainly not new; in the commutative case\footnote{Indeed, even in the noncommutative case it has
not gone entirely unnoticed. In \cite{Tong} a formal $2k$-parameter solution to
the vortex equations of the noncommutative Abelian-Higgs model was found to all
orders in $\gamma^{-1}$  and, in particular, the metric on the moduli space of
vortices explicitly computed in the limit $\gamma\rightarrow\infty$. There it
was also noted that while this limit is usually taken to mean
$\vartheta\rightarrow\infty$, it could equally well correspond to the large
coupling limit. It is this latter view that we advocate.}, this relation 
has been commented
on by several authors in many different contexts \cite{Hindmarsh92,
Hindmarsh93,Schroers,Witten93}. However, it remains to be seen whether this 
correspondence persists at the level of the solutions. If it does we will 
have produced an explicit descent from the vortices of the fuzzy linear 
sigma model to the instantons of the noncommutative $\C\Pk^{N}$ model. In the
interests of self-containment, we review now the derivation of the lump
solutions of the sigma model.\\
\begin{center}
\begin{figure}[h]{\epsfig{file=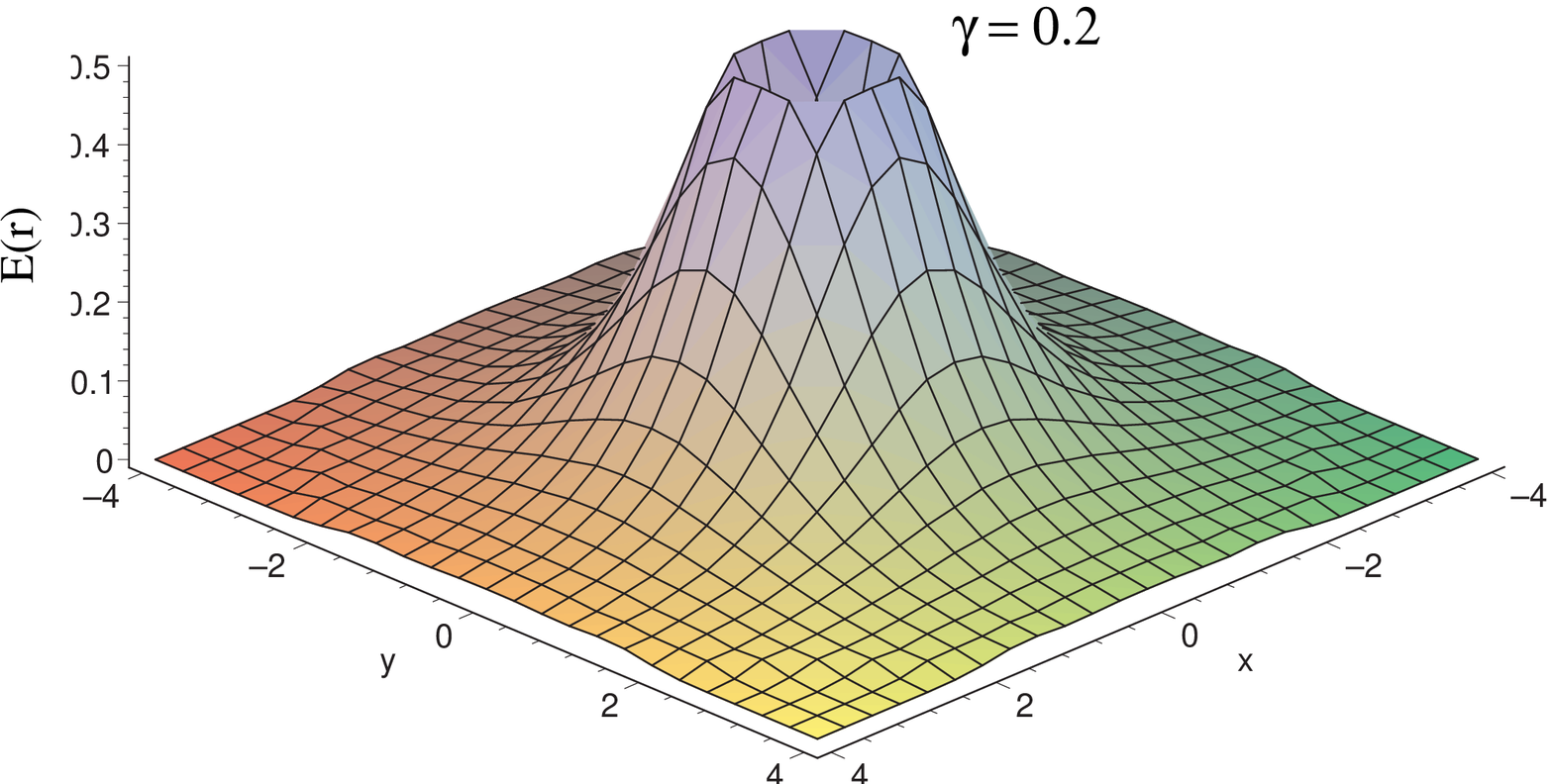,width=8cm,height=4.5cm}
        \epsfig{file=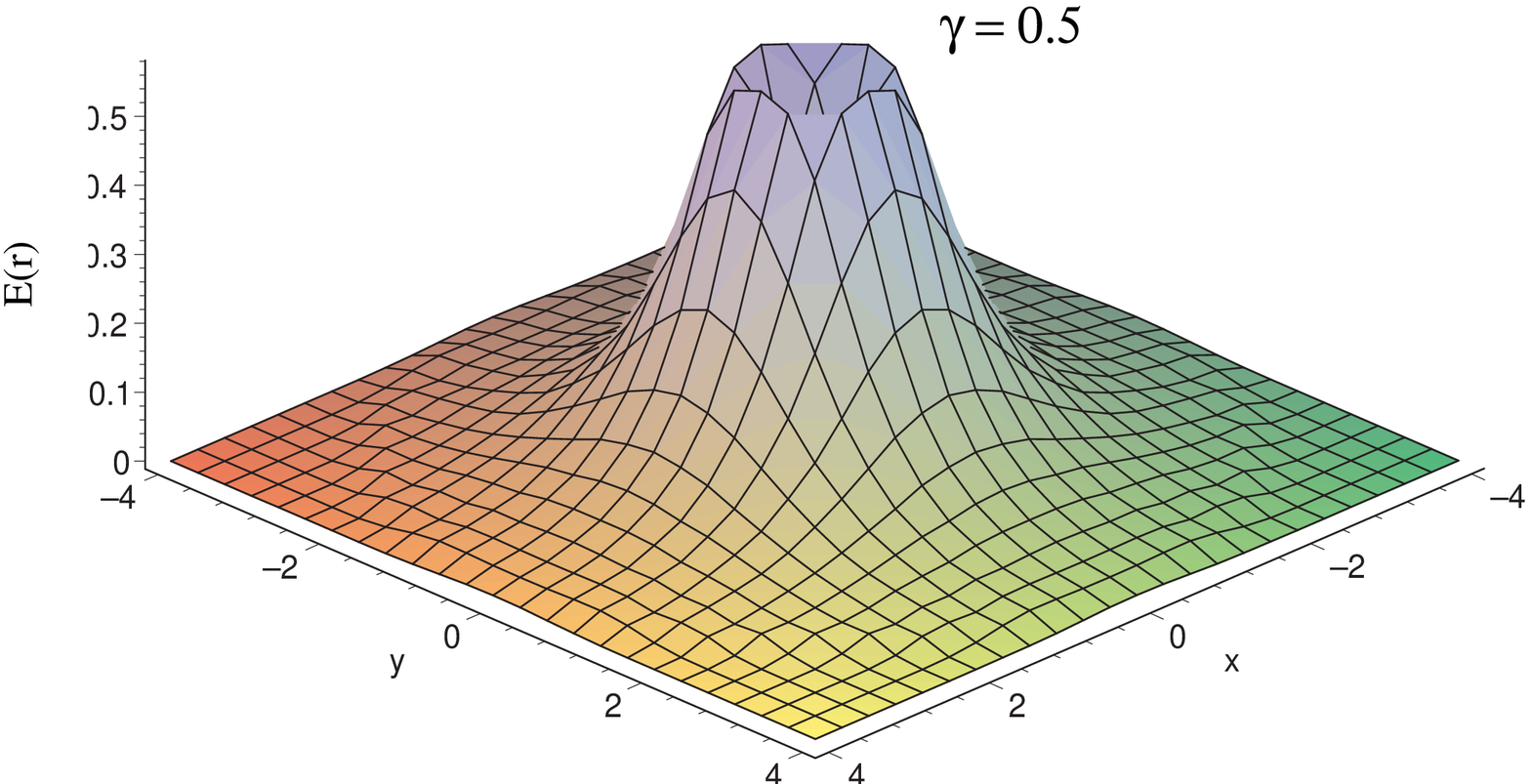,width=8cm,height=4.5cm}
	\epsfig{file=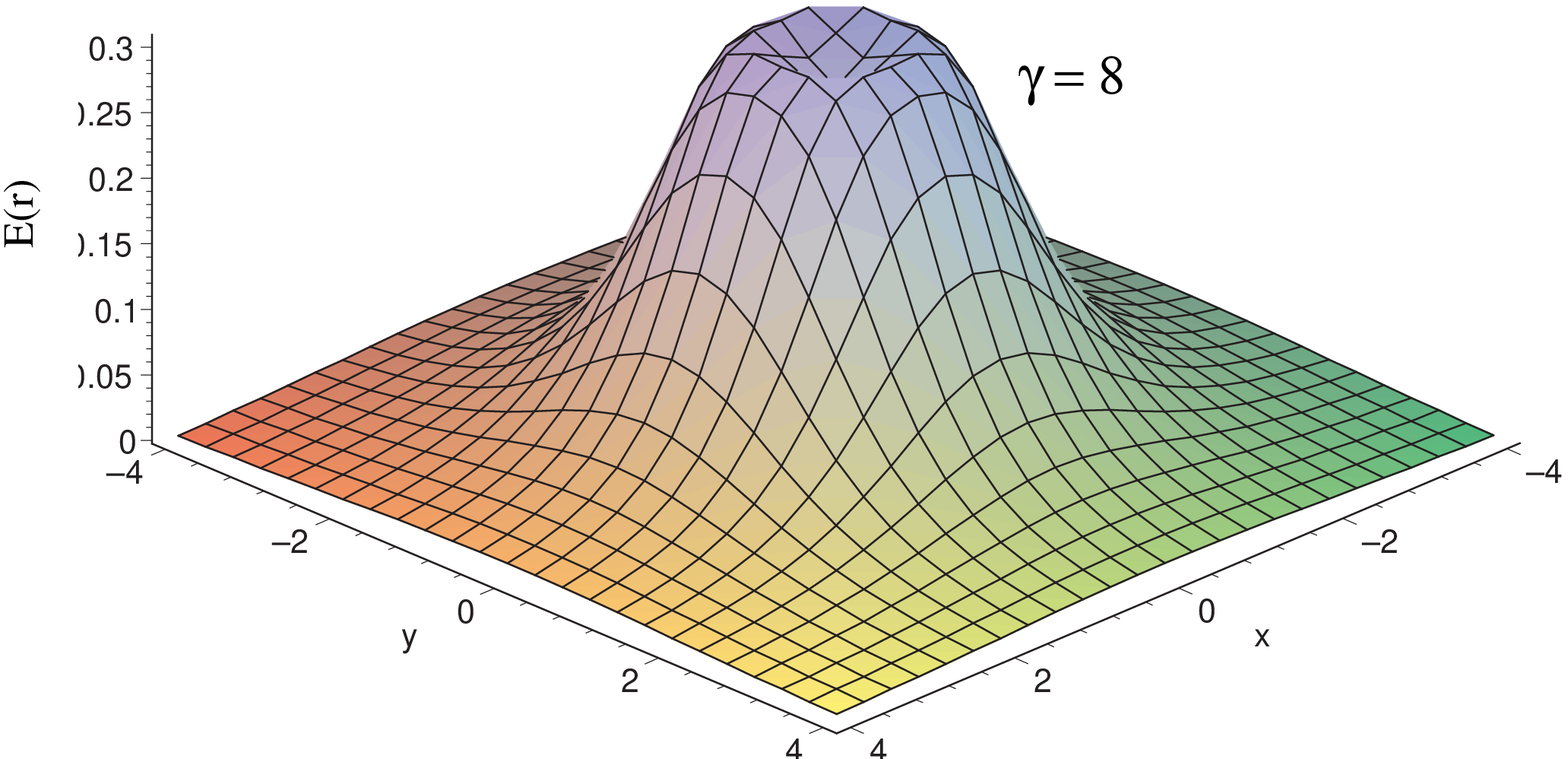,width=8cm,height=4.5cm}
	\epsfig{file=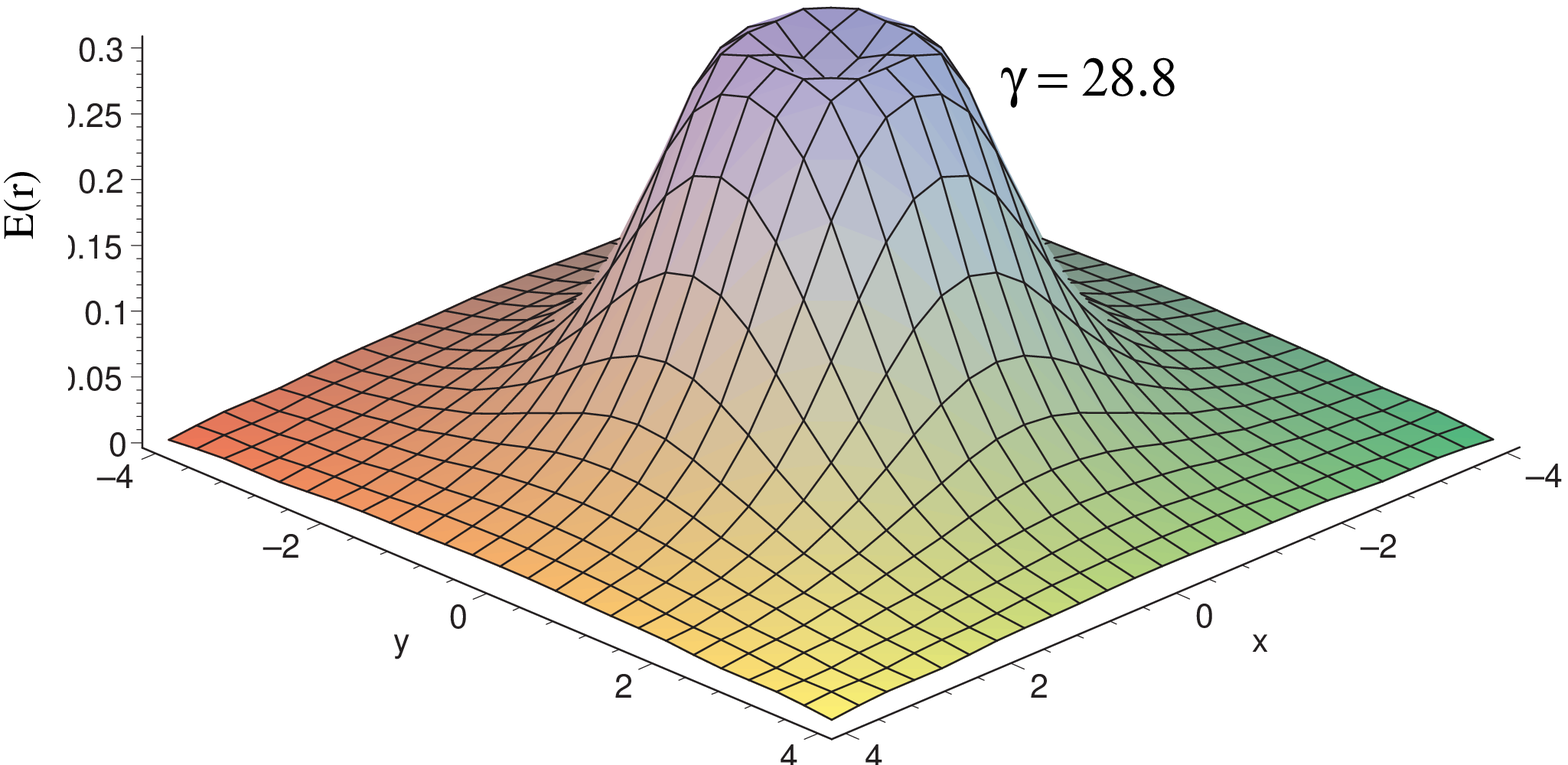,width=8cm,height=4.5cm}
 \caption{The metamorphosis of the semilocal vortex into the $\C\Pk^{N}$ lump}}
\end{figure}
\end{center} 
\noindent
With eq.(\ref{Large-Coupling-Energy}) as a starting point, a 
reparameterization of the $(N+1)-$component Higgs field as $\widehat{\Phi} =
(1/\sqrt{\widehat{W}\widehat{W}^{\dagger}})\widehat{W}$ and subsequent
definition of the Hermitian projector $P \equiv \widehat{W}^{\dagger}
(\widehat{W}\widehat{W}^{\dagger})^{-1}\widehat{W}$ allows for the static 
energy (or two-dimensional action) to be written as
\begin{eqnarray}
  E = 2\pi\,{\rm Tr}_{\cal H}{\rm
  tr}{\Bigl(}[P,\widehat{a}^{\dagger}][\widehat{a},P]{\Bigr)}.
  \label{NC-CPn-action}
\end{eqnarray}
In this form, the $\C\Pk^{N}$ energy is remarkably similar to the kinetic term of the
static energy of a $(2+1)-$dimensional noncommutative scalar field (see eq.(2.2)
of ref.\cite{GHS}) with the crucial difference of the additional matrix trace in
eq.(\ref{NC-CPn-action}). Indeed it was shown in \cite{LP2,LP3,Murugan-Adams}
that the quantity ${\rm Tr}_{\cal H}{\rm tr}
[\widehat{a},\widehat{a}^{\dagger}P]$ contributes a nonvanishing boundary 
term to the energy and some care needs to be exercised in the derivation 
of the noncommutative Bogomol'nyi bound. With this in mind, the energy may
correctly be written as 
\begin{eqnarray}
  E = 2\pi\,{\rm Tr}_{\cal H}{\rm tr}{\Bigl(}2F_{+}(P)^{\dagger}F_{+}(P){\Bigr)}
  + 2\pi Q_{+}\geq 2\pi Q_{+}
\end{eqnarray}   
with the topological charge $Q_{+}\equiv {\rm Tr}_{\cal H}{\rm tr}(P - 
[\widehat{a},\widehat{a}^{\dagger}P])$ and $F_{+}(P)\equiv (1-P)\widehat{a}P$.
A similar expression holds for the anti-BPS states.
Focusing on the BPS states though, saturation of the bound on the energy is 
obtained when $F_{+}(P) = 0$. As first shown in \cite{LeeLeeYang}, solutions 
are not difficult to find; any Hermitian projector constructed from an
$(n+1)-$vector $W$ whose components are holomorphic polynomials in $\widehat{z}$
will satisfy the above BPS equation. These are precisely the noncommutative
extension of the instanton solutions of the conventional $\C\Pk^{N}$ sigma model. 
For example, the static, $1-$ and $2-$lump solutions of the noncommutative 
$\CP$ model are given by
\begin{eqnarray}
  W_{1} = {\Bigl(}\widehat{z}-a_{1}\,,\,b_{1}{\Bigr)}\qquad 
  W_{2} = {\Bigl(}\widehat{z}^{2} - a_{2}\,,
  \,2b_{2}\widehat{z}+c_{2}{\Bigr)}
  \label{CP1-lumps}
\end{eqnarray}
where the soliton parameters $a_{1},...,d_{2}\in\C$ are chosen to coincide with the
standard way of writing the solutions in the commutative theory \cite{Ward85}. 
These are the
complex moduli of the $\CP$ instanton. To facilitate comparison with the
vortices, these may be written in the harmonic oscillator basis so that, for
example, the $1-$lump solution becomes
\begin{eqnarray}
 W_{1} =
 \sum_{n=0}^{\infty}\sqrt{\frac{\vartheta(n+1)}{\vartheta(n+1)+1}}|n\rangle\langle
 n+1|\otimes\langle {\rm I}| + \sum_{n=0}^{\infty}
 \sqrt{\frac{1}{\vartheta(n+1)+1}}|n\rangle\langle n|\otimes\langle {\rm I\!I}|.
 \label{CP1-lump-HObasis}
\end{eqnarray}
Returning to the degree$-k$ semilocal vortex of the last section, notice 
that eq.(\ref{Phi1-coefficients}) may be recast as
\begin{eqnarray}
  {\Bigl(}1+\frac{n!k!}{(n+k)!}\kappa^{2}{\Bigr)}Q_{n}-1 + 
  \frac{1}{\gamma}{\Bigl(}(n+k+1)\frac{Q_{n}}{Q_{n+1}} - 
  (n+k)\frac{Q_{n-1}}{Q_{n}} - 1{\Bigr)} = 0\quad{}
  \label{Expansion-Relations}
\end{eqnarray}
In the infinite coupling limit $e^{2}\rightarrow\infty$ (or equivalently
$\gamma\rightarrow\infty$), the above recurrence relation may be be solved
exactly to give
\begin{eqnarray}
  Q_{n} = {\Bigl(}1+\frac{n!k!}{(n+k)!}\kappa^{2}{\Bigr)}^{-1}.
  \label{Infinite-Coupling}
\end{eqnarray}
In particular, for $k=1$ we find
\begin{eqnarray}
  Q_{n} = \frac{n+1}{n+1+\kappa^{2}}\qquad P_{n} =
  \frac{\kappa^{2}}{n+1+\kappa^{2}}.
  \label{1-vortex-infinite-coupling}
\end{eqnarray}
Finally, matching coefficients to all orders in eqs.(\ref{CP1-lump-HObasis}) 
and (\ref{1-vortex-infinite-coupling}) means that the descent from
noncommutative vortex to fuzzy lump only occurs when $\kappa^{2} = 
1/\vartheta$. Indeed, this is exactly the choice we made in our numerical
computations to reduce the dimension of the shooting-parameter space. As a
check, we expect that for a fixed value of $\vartheta$, $Q_{0}\rightarrow
\vartheta/(\vartheta + 1)$ as $\gamma\rightarrow\infty$. 
A quick glance at the table of our numerical results verifies that this is 
indeed the case for $\vartheta = 0.2$ and $0.5$. Moreover, hindsight reveals that
the set of energy densities in figure 2. is in fact a series of snapshots of
the $k=1$ vortex of the noncommutative semilocal model morphing into a 
fuzzy $\CP$ $1-$lump. The case $k=2$ is no less straightforward. With its center
of mass localised at the origin, the $\CP$ $2-$lump in eq.(\ref{CP1-lumps}) 
can be written as
\begin{eqnarray}
  W_{2} &=&
 \sum_{n=0}^{\infty}\sqrt{\frac{\vartheta^{2}(n+1)(n+2)}{\vartheta^{2}(n+1)(n+2)+1}}
 |n\rangle\langle
 n+2|\otimes\langle {\rm I}|\qquad{}\nonumber\\
 &+& \sum_{n=0}^{\infty}
 \sqrt{\frac{1}{\vartheta^{2}(n+1)(n+2)+1}}|n\rangle\langle n|
 \otimes\langle {\rm I\!I}|
 \label{CP1-2lump-HObasis}
\end{eqnarray}  
when $b_{2}$, the frozen out modulus \cite{FurutaInami} is set to vanish. 
A comparison with the general expression for the infinite coupling coefficients 
(\ref{Infinite-Coupling}) reveals a matching at all levels only if $\kappa^{2} =
1/2\vartheta$. Generalisation to larger $k$ follows in much the same way so no 
further attention is paid to it here.\\

\noindent
At this juncture, a few comments are in order. The Bogomol'nyi equations of the
commutative gauged linear sigma model admit a one parameter family of vortex
solutions \cite{Hindmarsh93}. This single complex parameter $w$ is to the 
commutative theory what the ratio of initial coefficients $\kappa$ is to our
noncommutative model with $w=0$ corresponding to the conventional 
Neilsen-Olesen string. One of the distinguishing characteristics of the 
$w\neq0$ semilocal vortices is the power law behavior exhibited by the scalar
and gauge fields as they relax to their respective vacuum values. Consequently,
the magnetic field\footnote{Following \cite{Hindmarsh93} $\xi$ is a
dimensionless radial variable on the plane.} $B \sim 2|w|^{2}/\xi^{4}$ 
and the width of the flux tube trapped in the vortex core is an arbitrary 
parameter instead of the Compton wavelength of the vector boson as in the
Neilsen-Olesen vortex. In the noncommutative model we once again find a one
parameter family of vortices only now the parameter, $\kappa$, is not at all 
arbitrary. Indeed, we find that there exists a point in the $\kappa$ parameter
space dependent on the degree of the vortex and the deformation parameter
$\vartheta$ at which the semilocal vortex {\it
exactly} descends to the corresponding noncommutative $\C\Pk^{N}$ lump.
Correspondingly, the width of the magnetic flux tube associated with 
the semilocal vortex is set by the scale of noncommutativity. This observed
exact metamorphosis of the vortex into the lump should be compared to the
results of section 3. of \cite{Hindmarsh93}. There an expansion of
the $1-$instanton solution of the commutative $\C\Pk^{N}$ model in powers of 
$|w|/|z|$ was used to establish that the vortex-instanton matching was exact 
at spatial infinity with differences emerging at ${\cal O}(|w|^{4}/|z|^{4})$ in
this expansion.    
\section{Brane Realisations}
Quite apart from their intrinsic field theoretic value 
\cite{Vachaspati,Hindmarsh92,Hindmarsh93}, the vortices of gauged linear sigma
models also have a remarkably rich stringy structure. Beginning with the 
ground-breaking work of \cite{Hanany-Witten} in which the 
$(2+1)-$dimensional, ${\cal N}=4$ $U(N)$ Yang-Mills-Higgs theory was recognised 
as the worldvolume theory on
a stack of $N$ $D3-$branes suspended between two parallel $NS5-$branes, an 
intricate tapestry of ideas can be woven, leading inexorably to a realisation 
of the noncommutative semilocal vortex as a $D-$brane configuration in type IIB
string theory \cite{Hanany-Tong}. In this section, we review some of these ideas
and cast them into a form that better facilitates comparison with our results.\\

\noindent
As in \cite{Hanany-Tong} the description of the system begins with a
$(2+1)-$dimensional, ${\cal N}=4$, $U(N)$ Yang-Mills-Higgs theory. The field
content of the theory consists of a $U(N)$ vector multiplet made up of a
gauge field $A_{\mu}$ and a triplet of adjoint scalars $\phi^{r}$ together with
their fermionic super partners. Coupled to these are $N$ fundamental
hypermultiplets each of which contain a doublet of complex scalars $q$ and
$\widetilde{q}$ and their super partners. The Lagrangian for the theory is
endowed with a global $SU(N+M)$ flavour symmetry as well as a local $U(N)$ gauge
symmetry. Consequently, under these two groups and with $N_{f}\equiv N+M$
denoting the number of flavours, $q$ and $\widetilde{q}$ 
transform as $({\bf N},\overline{\bf N_{f}})$ and $(\overline{\bf N},
{\bf N_{f}})$ respectively; the fundamental scalars are represented by
$N\times(N+M)$ matrices. The dynamical content of the bosonic sector of the
theory is contained in the Lagrangian 
\begin{eqnarray}
  {\cal L} &=& -{\rm Tr}
  {\Bigl[}\frac{1}{4e^{2}}F^{2} + \frac{1}{2e^{2}}(D\phi^{r})^{2} 
  + (Dq)^{2} + (D\widetilde{q})^{2} + e^{2}|q\widetilde{q}|^{2}\nonumber\\
  &+& \frac{1}{2e^{2}}[\phi^{r},\phi^{s}]^{2} + (q^{2}+\widetilde{q}^{2})
  \phi^{r}\phi^{r} + \frac{e^{2}}{2}(q^{2}-\widetilde{q}^{2} - 
  \zeta{\bf 1}_{N}){\Bigr]}.
  \label{SYMH-action}
\end{eqnarray}
where the Fayet-Illiopolous (FI) parameter, $\zeta$, in the final D-term in 
(\ref{SYMH-action}) is chosen to be positive. This theory exhibits a Higgs
branch of vacua which possess BPS vortices only if 
$\widetilde{q}$ and $\phi^{r}$ both vanish. This
constraint defines a so-called reduced Higgs branch, ${\cal
N}_{N,M}=Gr(N,N+M)$, the Grassmannian manifold of $N-$dimensional
hyperplanes in $\C^{N+M}$. A particular vacuum choice\footnote{Since the
Grassmannian is, after all, a symmetric space, no generality is lost in this
choice.} is made by picking
\begin{eqnarray}
  q_{\rm vac} &=& \left\{ \begin{array}{ll}
        \sqrt{\zeta}\delta^{a}_{i} & a,i=1,...,N\\
        0 & i=N+1,...,N+M
	\end{array}\right.  	 
\end{eqnarray}
\FIGURE{\epsfig{file=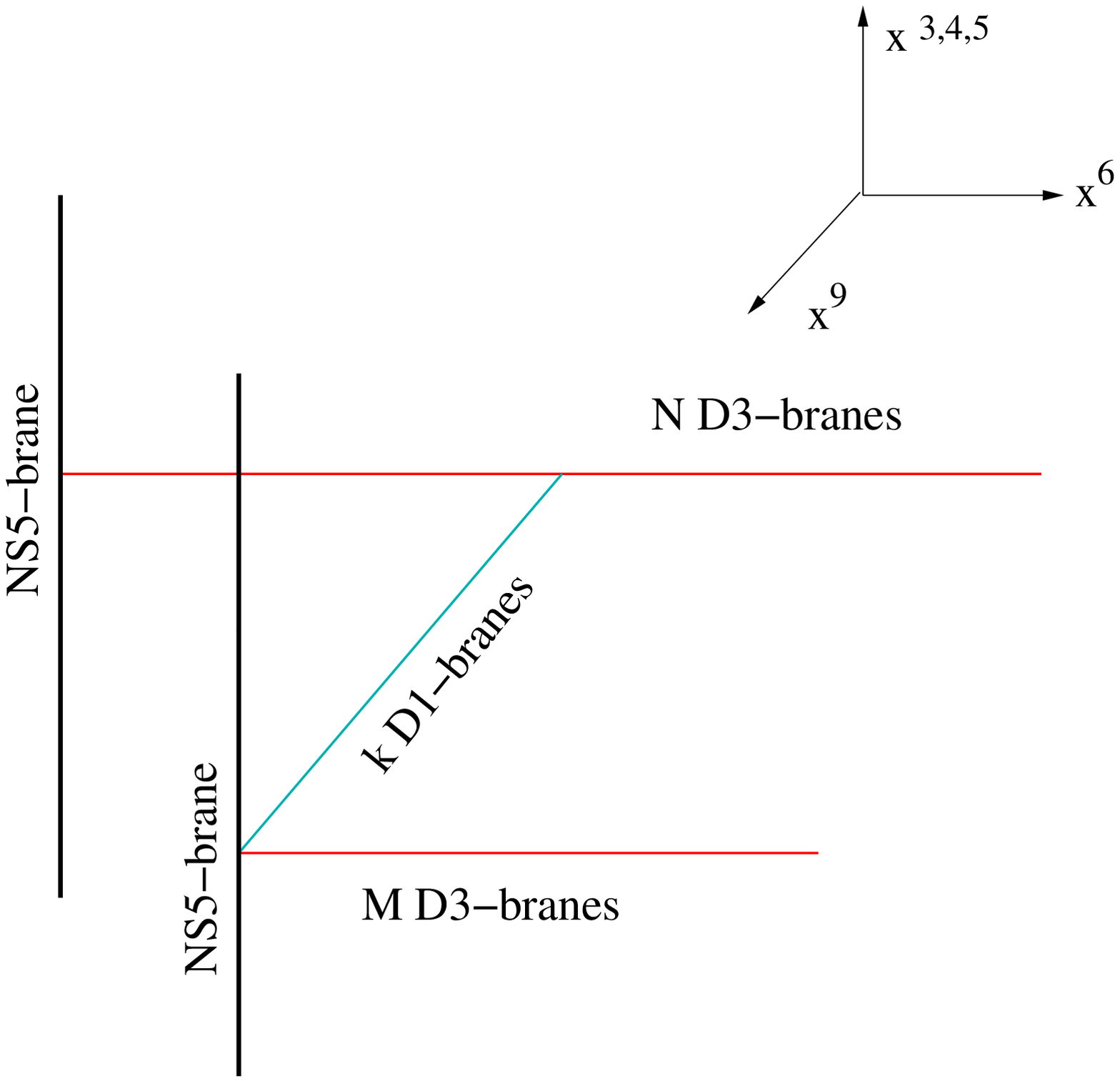,width=6cm,height=6cm}         
\caption{The degree$-k$ BPS vortex as $k$ stretched D-strings.}}
\noindent
In our abelian case, for example, $N = 1$, the reduced Higgs branch
${\cal N}_{1,M} = Gr(1,1+M) \simeq \C\Pk^{M}$ and $q_{\rm vac} =
(\sqrt{\zeta},0)$. Relabeling $q\rightarrow\Phi$, setting the FI
parameter $\zeta = 1$ and restricting to time-independent solutions 
trivially establishes the equivalence of the action in this branch with 
the static energy (\ref{StaticEnergy3}). As discussed earlier, the
spectrum of solutions of this theory is rich with BPS vortices. The brane
realisation of these vortices is built up from the $U(N)$ 
Yang-Mills-Higgs described in (\ref{SYMH-action}). It consists of $N$ 
$D3-$branes suspended between two parallel $NS5-$branes and a further 
$N+M$ $D3$'s attached to the right hand $NS5-$brane to add flavour (see
figure 3).\\ 

\noindent
In the
Higgs branch, one of the $NS5-$branes is separated from the others. This
separation is proportional to the FI parameter $\zeta$. The degree$-k$ BPS 
vortices manifest as $k$ $D-$strings stretched between the
$D3-$branes and the separated $NS5-$brane - an identification made on
the basis of the fact that the stretched $D1-$branes are the {\it only} BPS
states of the brane configuration with the correct mass.
More than just a 
pretty picture, the geometry of the $D-$brane 
configuration in figure 3 encodes vital information about the FI parameter,
$\zeta$ as well as the gauge coupling $e$ as
\begin{eqnarray}
  \frac{1}{e^2}=\frac{\Delta x^{6}}{2\pi g_{s}};\quad 
  \zeta = \frac{\Delta x^{9}}{4\pi^{2}g_{s}l_{s}^{2}}
  \label{FI-Coupling-relation}
\end{eqnarray}
where $l_{s}$ and $g_{s}$ are the string length and coupling
respectively and $\Delta x^{6}$ and $\Delta x^{9}$ are the separation
distances between the $NS5-$branes defined as in figure 1. 
It is now clear that the sigma model limit ($e^{2}\rightarrow\infty$) 
of the vortex occurs precisely when the separation of the $NS5-$branes 
in the $6-$direction vanishes. The configuration that realises the
$k-$lump solution of the (commutative) $\C\Pk^{M}$ nonlinear sigma model
then is as above only with $N=1$.
In string theory the transition from commutative to noncommutative 
worldvolume theories is achieved by turning on an NS-NS $B-$field in the
appropriate direction \cite{SeibergWitten}. In the present context, the
transition from the semilocal action (\ref{StaticEnergy3}) to its
noncommutative counterpart (\ref{NCStaticEnergy1}) translates into turning
on a constant NS-NS $B-$field $B_{12}=\vartheta\,dx^{1}\wedge dx^{2}$ in the 
$(1,2)-$directions in a background of two $NS5-$branes with a $D3-$brane
stretched between them and a further $M+1$ $D3$'s attached to the right 
hand $NS5-$brane. What of the vortices? The effect of the $B-$field on the
$D-$strings stretched between the $NS5-$brane and the $D3$ 
is quite remarkable. The basic physics is analogous to the
situation of a $D-$string suspended between two $D3-$branes studied in
\cite{Hashimoto} and was first described for the vortex case 
in \cite{Hanany-Tong}. The NS-NS
$2-$form manifests on the $D3-$worldvolume as a constant magnetic flux
${\cal F}_{12}$ while the $D-$string endpoint appears as a magnetic source.
Since on the $4-$dimensional worldvolume of the $D3-$brane ${\cal F}_{12} =
\star {\cal F}_{06}$, the magnetic endpoint of the $D1-$brane feels the same
force as an electric charge in a constant electric field in the  
$6-$direction. However, as other end of the $D-$string remains married to
the $NS5-$brane, the $D-$string responds to this force by tilting as in
figure 4. The effect of the tilting was investigated in \cite{Hashimoto} by
studying the $D-$string Born-Infeld action at weak string coupling
\begin{eqnarray}
  S = \frac{1}{2\pi l_{s}^{2}}\int_{0}^{\Delta x^{9}}\!\!dx^{9}\,
  {\Bigg(}\frac{1}{g_{s}}\sqrt{1+{\Bigl(}\frac{dx^{6}}{dx^{9}}
  {\Bigr)}^{2}} + A_{06}\frac{dx^{6}}{dx^{9}}{\Bigg)}
  \label{BIEnergy} 
\end{eqnarray}
where the RR $2-$form $A_{06}$ that couples to the $D-$string worldvolume is 
induced by ${\cal B}_{06}$. The result of that investigation translated into
the language of the vortex theory \cite{Hanany-Tong} is that the 
displacement of the $D1-$brane endpoint is given 
by\footnote{Note the sign difference from \cite{Hanany-Tong} and the
difference it has on the tilt of the $D-$strings.} 
$\delta = (\vartheta 
\Delta x^{9})/(2\pi l_{s}^{2})$. With this and some straightforward algebra,
the distance between the $D-$string endpoint and the left $NS5-$brane can be
computed.  With the choice of $\zeta = 1$ for the FI parameter, 
the result is 
\begin{eqnarray}
  r = 2\pi g_{s}{\Bigl(}\frac{1}{e^{2}} + \vartheta{\Bigr)}.
  \label{Vortex-FI-parameter}
\end{eqnarray}
\FIGURE{\epsfig{file=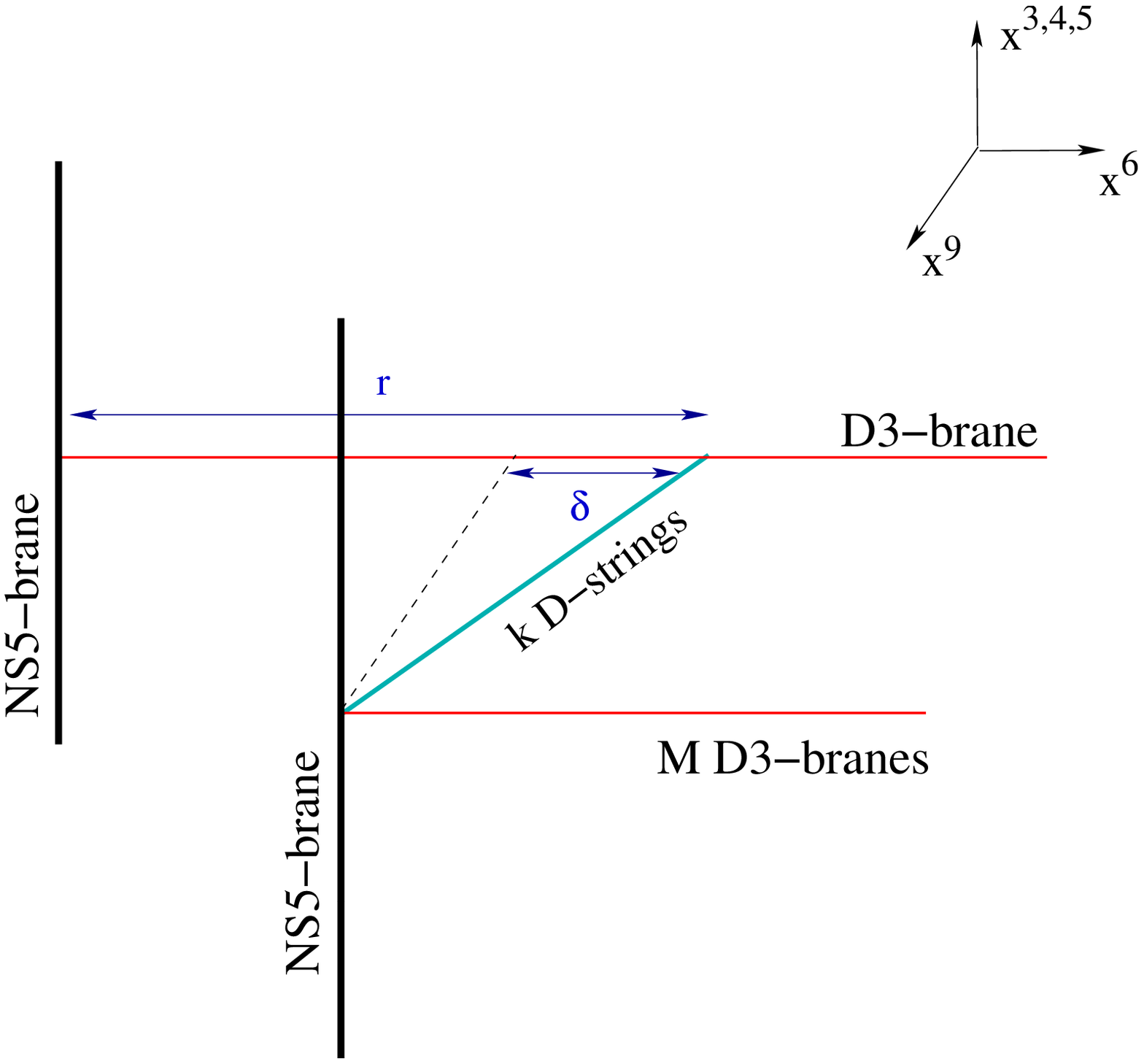,width=6cm,height=6cm}         
\caption{The magnetic field induced on the $D3-$worldvolume causes the
$D-$strings to tilt.}}
\noindent 
This distance is, in fact, the FI parameter of the theory living on the 
$D1-$branes (see \cite{Hanany-Tong} for a lucid discussion of this aspect).
Having fixed $\Delta x^{9}$
with the choice $\zeta=1$ the magnitude of $r$ is completely determined by
the size of the gauge coupling as determined by the $NS5-$brane 
separation in the $6-$direction and the noncommutativity. Since the latter
is also fixed, the transition from vortex to lump can be studied by
changing the distance between the $NS5-$branes. As $\Delta x^{6}$ is
decreased to zero, the separation between the
$D-$string endpoint and the left $NS5-$brane decreases to 
$r_{*} = 2\pi g_{s}\vartheta$. It is this configuration of the $k$ tilted
$D-$strings stretched between the (formerly right hand) $NS5-$brane 
and the $D3-$brane that realises the degree$-k$ instanton of the 
$\C\Pk^{M}$ sigma model. This concludes our treatment of $D-$brane realisation
of the noncommutative $\C\Pk^{M}$ lump.\\

\noindent
More than just an academic exercise, 
this identification of the semilocal vortex
and $\C\Pk^{M}$ instanton has proven invaluable in the understanding of the low
energy dynamics of both the vortex and instanton as encoded in the geometry of
their respective moduli spaces \cite{Tong}. We refer the interested reader 
to \cite{Hanany-Tong} for a nice discussion of the structure of the moduli
spaces and content ourselves with merely summarising some of their most 
pertinent results. The moduli space of degree$-k$ semilocal vortices 
$\hat{{\cal V}}_{k,(1,M)}$ is a $2k(1+M)-$dimensional space with a 
natural K\"ahler metric defined by the overlap of zero modes. However, 
this metric is afflicted with some non-normalisable zero modes that, 
classically, correspond to the moduli with infinite moments of inertia 
and that make the quantum mechanical treatment of these solitonic objects 
quite subtle. Fortunately these subtleties may be circumvented with a 
little help from the branes. A study of the theory on the $D1-$brane
predicts that the Higgs branch, $\hat{\cal M}_{k,(1,M)}$, constructed by a 
$U(k)$ K\"ahler quotient of $\C^{k(1+M+k)}$ is isomorphic to the moduli
space $\hat{{\cal V}}_{k,(1,M)}$. While the metric on 
$\hat{\cal M}_{k,(1,M)}$ retains all the symmetries of the K\"ahler metric
on the vortex moduli space, it is finite and suffers from none of the 
non-normalisablity problems of the latter. Consequently, the study of the
quantum theory of semilocal vortices may be simplified somewhat by 
replacing the natural metric on the vortex moduli space  
with the metric on the Higgs branch of the $D-$string theory inherited 
from the K\"ahler quotient construction of \cite{Hanany-Tong}.  
       
\section{Conclusions and Discussion}
\noindent
The primary concern of this work has been the construction and study
of a noncommutative extension of $(2+1)-$dimensional critically coupled, 
gauged linear sigma model. Like its commutative counterpart this theory
possesses a rich spectrum of BPS solutions. By extending the
systematic construction of \cite{Lozano1} we have explicitly constructed 
a family of vortex solutions to the BPS equations 
(\ref{NCBPS-equations}) for arbitrary positive values 
of the noncommutativity parameter $\vartheta$. As expected, these
fuzzy vortices reduce to the exact Neilsen-Olesen strings of the
noncommutative Abelian-Higgs model \cite{DongsuBak,Bak,Lozano1,Wadia} 
on the co-dimension one surface $\kappa=0$ of the parameter space. 
Despite retaining many of the properties of their commutative cousins
\cite{Hindmarsh93,Vachaspati}, the introduction of a new length scale set 
by the noncommutativity parameter $\vartheta$ induces several remarkable
differences. Among these we find that the width of the magnetic flux tube 
trapped in the vortex core no longer exhibits the characteristic
arbitrariness of the commutative semilocal vortex. In the noncommutative
model, this width is set by the scale of the noncommutativity.\\

\noindent
The detailed investigation of the large coupling 
($e^{2}\rightarrow\infty$) regime of the $\vartheta-$deformed gauged 
linear sigma model carried out in section 4. confirms, both numerically and
analytically, the 
commutative intuition of the vortex morphing into a lump of the (fuzzy)
$\C\Pk^{M}$ sigma model. Additionally, while the agreement between vortex
and lump in the $\vartheta = 0$ case is precise only asymptotically
\cite{Hindmarsh93}, we find an {\it exact} matching at all levels of 
the harmonic oscillator expansion at finite $\vartheta$. Indeed, insisting
that this agreement holds selects a preferred set of values for $\kappa$,
dependent on the scale of noncommutativity and the degree of the vortex.
This effectively reduces the dimension of the parameter space by one. While
we have explicitly constructed solutions for the $1-$ and $2-$vortex cases,
the construction of higher degree solutions follows in much the same way and
we do not expect any further surprises.\\

\noindent
Finally, we reviewed the elegant constructions of \cite{Hanany-Tong} that
lead to a realisation of the noncommutative $\C\Pk^{M}$ $k-$lump as 
$k$ tilted $D-$strings stretched between an isolated $NS5-$brane (on which a
stack of $M$ semi-infinite $D3-$branes end) and a
semi-infinite $D3$ whose one endpoint ends on a second $NS5$ 
(see figure 4). This identification is built on the foundation of a study
of the ${\cal N} = 4$ $U(N)$ Yang-Mills-Higgs $D3-$worldvolume theory 
hinges on the metamorphosis of vortices into lumps. Of course, to
be sure that this configuration really does correspond to the lump
solution requires more work than just a comparison of the masses of both
configurations; the spectrum of fluctuations around each
object needs to be computed and compared. This is a more difficult endeavor
which, together with a more thorough investigation of the spectrum of BPS
objects of the noncommutative gauged linear sigma model is left to future 
work \cite{Murugan-Millner2}. Curiously, this realisation 
of fuzzy $\C\Pk^{M}$ lumps is not unique, at least for $M=1$. 
Drawing on the tree level equivalence between ${\cal N} = 2$ open string 
theory and self-dual Yang-Mills theory in $(2+2)-$dimensions \cite{OV}, 
it was argued in \cite{LP2,LP3} that the effective field theory induced 
on the worldvolume of $N$ $D2-$branes by ${\cal N}=2$ open strings in a 
K\"ahler $B-$field background is a noncommutative $U(N)$ sigma model. 
Using a modified ``method of dressing" soliton solutions of the latter 
were constructed and their various scattering properties investigated. 
In this context, the $k-$lump solution of the $\C\Pk^{1}$ sigma model 
may be interpreted as $k$ $D0-$branes in the worldvolume of a stack of 
$D2-$branes \cite{LP3,Ihl}. Again, while this assertion needs to be tested 
beyond the level of a mass comparison, the possibility of a duality between
${\cal N} = 2$ open string theory and the type II-B superstring is, to say
the least, intriguing and certainly deserves more attention.

\acknowledgments
 J.M. would like to thank the KITP (Santa Barbara) for the warm hospitality and 
 a most stimulating environment during the early stages of the work. 
 Our gratitude is extended to Robert de Mello Koch, George Ellis, 
 Clifford Johnson, David Tong and Amanda Weltman for valuable discussions at 
 various stages of this work and to Bill Watersson who inspired the title. 
 And finally, a word from our sponsors: J.M. is supported by the 
 Sainsbury-Lindbury Trust and a research associateship of the University 
 of Cape Town. A.M acknowledges the NRF and the University of Cape Town for
 financial support.

{\small\begin{thebibliography}{100}

\bibitem{DongsuBak}
D. Bak,
``{\it Exact Multivortex Solutions in Noncommutative Abelian Higgs Theory}''
{\it Phys. Lett. B} {\bf 495}, 251-255, (2000), \texttt{hep-th/0008204}
 
\bibitem{Bak}
D. Bak, K. Lee and J-H. Park,
``{\it Noncommutative Vortex Solitons}''
{\it Phys. Rev. D} {\bf 63}, 125010 (2001), \texttt{hep-th/0011099}

\bibitem{deVega}
H. de Vega and F. Schaposnik,
``{\it A classical vortex solution of the of the Abelian-Higgs model}''
{\it Phys. Rev. D} {\bf 14}, 1100 (1976)

\bibitem{Douglas-Nekrasov}
M.R. Douglas and N.A. Nekrasov ,
``{\it Noncommutative Field Theory}''
{\it Rev. Mod. Phys} {\bf 73}, 977 (2001), \texttt{hep-th/0106048}

\bibitem{FurutaInami} 
K. Furuta, T. Inami, H. Nakajima and M. Yamamoto, 
``{\it Low-energy dynamics of noncommutative $\CP$ solitons in 
$(2+1)-$dimensions,}" {\it Phys. Lett. B} {\bf 537}, 165-172, 
(2002), \texttt{hep-th/0203125} 

\bibitem{FurutaInami2} 
K. Furuta, T. Inami, H. Nakajima and M. Yamamoto, 
``{\it Non-BPS Solutions of the Noncommutative $\CP$ Model in 
$(2+1)$-dimensions,}" {\it JHEP} {\bf 0208}, 009, 
(2002), \texttt{hep-th/0207166} 

\bibitem{GMS}
R. Gopakumar , S. Minwalla and A. Strominger,
``{\it Noncommutative Solitons}"
{\it JHEP} {\bf 0005}, 020 (2000), \texttt{hep-th/0003160}

\bibitem{GHS}
R. Gopakumar, M. Headrick and M. Spradlin,
``{\it On Noncommutative Multisolitons.}" 
(2001), \texttt{hep-th/0103256}

\bibitem{Gross}
D.J. Gross and N.A. Nekrasov,
``{\it Solitons in noncommutative gauge theory.}"
{\it JHEP} {\bf 0103}, 044 (2001), \texttt{hep-th/0010090}
 
\bibitem{Hanany-Tong}
A. Hanany and D. Tong,
``{\it Vortices, Instantons and Branes}"
\texttt{hep-th/0306150}

\bibitem{Hanany-Witten}
A. Hanany and E. Witten,
``{\it Type IIB superstrings, BPS monopoles, and three-dimensional gauge
invariance}"
{\it Nucl. Phys. B} {\bf 492}, 152 (1997)
\texttt{hep-th/9611230}

\bibitem{Harvey1} J.A. Harvey, 
``{\it Komaba Lectures on Noncommutative Solitons and D-Branes,}" (2001), 
\texttt{hep-th/0102076} 

\bibitem{HarveyKrausLarsenMartinec} J.A. Harvey, P. Kraus, F. Larsen 
and E. J. Martinec, 
``{\it D-Branes and Strings as Non-commutative solitons,}" JHEP {\bf 0007}, 042
(2000), 
\texttt{hep-th/0005031}

\bibitem{HarveyKrausLarsen} J.A. Harvey, P. Kraus and F. Larsen, 
``{\it Exact Noncommutative Solitons,}" JHEP {\bf 0012}, 024
(2000), 
\texttt{hep-th/0010060}

\bibitem{Hashimoto} A. Hashimoto and K. Hashimoto,
``{\it Monopoles and dyons in noncommutative geometry}" JHEP {\bf 11}, 005
(1999)
\texttt{hep-th/9909202}

\bibitem{Hindmarsh92}
M. Hindmarsh,
{\it ``Existence and Stability of Semilocal Strings"}
{\it Phys. Rev. Lett.} {\bf 68}, 9 (1992)

\bibitem{Hindmarsh93}
M. Hindmarsh,
{\it ``Semilocal Topological Defects"}
{\it Nucl. Phys. B} {\bf 392}, 461 (1993)
\texttt{hep-ph/9206229}

\bibitem{Ihl} M. Ihl and S. Uhlmann,
``{\it Noncommutative Extended Waves and Soliton-like Configurations in N=2 
String Theory,}'' (2002),
\texttt{hep-th/0211263}

\bibitem{LP1}
O.Lechtenfeld and A.D.Popov,
``{\it Noncommutative Multi-solitons in 2+1 dimensions}"
{\it JHEP} {\bf 0111}, 040 (2001), \texttt{hep-th/0106213}

\bibitem{LP2}
O.Lechtenfeld and A.D.Popov,
``{\it Scattering of Noncommutative Solitons in 2+1 dimensions}"
{\it Phys. Lett. B} {\bf 523}, 178-184 (2001), \texttt{hep-th/0108118}

\bibitem{LP3}
O.Lechtenfeld, A.D.Popov and B. Spendig,
``{\it Noncommutative Solitons in Open N=2 String Theory}"
{\it JHEP} {\bf 0106}, 011 (2001), \texttt{hep-th/0103196}

\bibitem{LeeLeeYang}
B-H. Lee, K. Lee and H.S. Yang,
``{\it The $\C\Pk^{N}$ model on noncommutative plane}"
{\it Phys. Lett B} {\bf 498}, 277-284 (2001), \texttt{hep-th/0007140}

\bibitem{Lozano1}
G.S. Lozano, E.F. Moreno and F.A. Schaposnik,
``{\it Nielsen-Olesen vortices in noncommutative space}"
{\it Phys. Lett B} {\bf 504}, 117 (2001), \texttt{hep-th/0011205}

\bibitem{Manton82}
N.S. Manton,
``{\it A remark on the scattering of BPS monopoles}"
{\it Phys. Lett B} {\bf 110}, 54 (1982)

\bibitem{Murugan-Adams}
J. Murugan and R. Adams,
``{\it Comments on Noncommutative Sigma Models}"
{\it JHEP} {\bf 0212}, 073 (2002), \texttt{hep-th/0211171}

\bibitem{Murugan-Millner2}
J. Murugan and A. Millner,
{\it In progress}

\bibitem{Nekrasov1}
N.A. Nekrasov,
``{\it Noncommutative instantons revisited.}" 
(2000), \texttt{hep-th/0010017}

\bibitem{Nielsen-Olesen}
H.B. Nielsen and P. Olesen,
``{\it Vortex line models for dual strings}"
{\it Nucl. Phys. B} {\bf 61}, 45 (1973)

\bibitem{OV}
H. Ooguri and C. Vafa,
``{\it Geometry of N=2 Strings}''
{\it Nucl. Phys. B} {\bf 361}, 469-518 (1991)

\bibitem{Schroers}
B.J. Schroers,
``{\it The Spectrum of Bogomol'nyi solitons in gauged linear sigma models}''
{\it Nucl. Phys. B} {\bf 475}, 440-468 (1996), \texttt{hep-th/0210010}

\bibitem{SeibergWitten}
N. Seiberg and E. Witten,
``{\it String Theory and Noncommutative Geometry}"
{\it JHEP} {\bf 09}, 032 (1999), \texttt{hep-th/9908142}

\bibitem{Strachan}
I. Strachan,
``{\it Low velocity scattering of vortices in a modified Abelian-Higgs model}"
{\it J. Math. Phys.} {\bf 33}, 102 (1992)

\bibitem{Szabo}
R.J. Szabo,
{\it ``Quantum Field Theory on Noncommutative Spaces"}
{\it Phys. Rep.} {\bf 378}, 207 (2003)

\bibitem{Tong}
D. Tong,
``{\it The Moduli Space of Noncommutative Vortices}"
{\it J. Math. Phys.} {\bf 44}, 3509 (2003)
\texttt{hep-th/0210010}

\bibitem{Vachaspati}
T. Vachaspati and A. Achucarro,
``{\it Semilocal cosmic strings}"
{\it Phys. Rev. D} {\bf 44}, 3067 (1991)

\bibitem{Ward85}
R.S. Ward,
``{\it Slowly moving lumps in the $CP(1)$ model in $(2+1)-$dimensions}"
{\it Phys. Lett B} {\bf 158}, 424 (1985)

\bibitem{Wadia}
D. Jatkar, G. Mandal and S. Wadia,
``{\it Nielsen-Olesen Vortices in Noncommutative Abelian Higgs Model}"
{\it JHEP} {\bf 0009}, 018 (2000), \texttt{hep-th/0007078}

\bibitem{Witten77}
E. Witten,
``{\it Some exact multi-instanton solutions of classical Yang-Mills theory}"
{\it Phys. Rev. Lett.} {\bf 38}, 121 (1977)

\bibitem{Witten86}
E. Witten,
``{\it Noncommutative Geometry and String Field Theory}"
{\it Nucl. Phys. B} {\bf 268}, 253 (1986)

\bibitem{Witten93}
E. Witten,
``{\it Phases of $N=2$ theories in two dimensions}"
{\it Nucl. Phys. B} {\bf 403}, 159 (1993)

\end{thebibliography}}
\end{document}